\documentclass{aa}  
\usepackage{aas_macros}
\usepackage{graphicx}
\usepackage{multirow}
\usepackage{amsmath}	
\usepackage{color, soul}
\usepackage{ulem,xcolor,tikz}

\usepackage{tabularx}

\usepackage{txfonts}
\begin{document}

   \title{HD\,163296 and its Giant Planets: Creation of Exo-comets, Interstellar Objects and Transport of Volatile Material}

   \subtitle{}
   \titlerunning{HD\,163296: Transport of Volatile Material}
   
   \author{D. Polychroni\thanks{Corresponding author}\inst{1,2,3},
          D. Turrini\inst{1,3}, S. Ivanovski\inst{2}, F. Marzari\inst{4}, 
          L. Testi\inst{5}, R. Politi\inst{6,3},  A. Sozzetti\inst{1}, 
          J. M. Trigo-Rodriguez\inst{7,8}, S. Desidera\inst{9},
          M. N. Drozdovskaya\inst{10}, S. Fonte\inst{6}, S. Molinari\inst{6}, 
          L. Naponiello\inst{1}, E. Pacetti\inst{6}, E. Schisano\inst{6}, 
          P. Simonetti\inst{2,3} \and M. Zusi\inst{6}
          }
   \authorrunning{D. Polychroni et al.}
   \institute{INAF -- Osservatorio Astrofisico di Torino, 
              via Osservatorio 20, I-10025, Pino Torinese, Italy\\
              \email{danae.polychroni@inaf.it}
         \and INAF -- Osservatorio Astronomico di Trieste,
              via G. B. Tiepolo 11,I-34143, Trieste, Italy
         \and ICSC – National Research Centre for High Performance Computing,  
              Big Data and Quantum Computing, Via Magnanelli 2, 40033,
              Casalecchio di Reno, Italy
         \and Dipartimento di Fisica e Astronomia – Università di Padova, 
              via Marzolo 8, 35121 Padova, Italy
         \and  Dipartimento di Fisica e Astronomia, Università di Bologna,
               via Gobetti 93/2, 40122, Bologna, Italy
         \and INAF – Istituto di Astrofisica e Planetologia Spaziali, 
              via Fosso del Cavaliere 100, 00133, Roma, Italy
         \and Institut de Ciències de l’Espai (ICE-CSIC), 
              Campus UAB, 08193 Cerdanyola del Vallès, Catalonia, Spain
         \and Institut d’Estudis Espacials de Catalunya (IEEC), Esteve Terradas 1, Edifici RDIT, Oficina 212, 
              PMT, Campus UPC, 08860 Castelldefels (Barcelona), Catalonia, Spain
         \and INAF--Osservatorio Astronomico di Padova, 
              Vicolo Osservatorio 5, 35122, Padova, Italy
         \and Physikalish-Meteorologisches Observatorium Davos und          
              Weltstrahlungszentrum (PMOD/WRC), Dorfstrasse 33, 7260, 
              Davos Dorf, Switzerland
          }

   \date{Received November 20, 2024; accepted March 6, 2025}
 
  \abstract
    {
    The birth of giant planets in protoplanetary disks is known to alter the structure and evolution of the disk environment, but most of our knowledge focuses on its effects on the observable gas and dust. The impact on the evolution of the invisible planetesimal population is still limitedly studied, yet mounting evidence from the Solar System shows how the appearance of its giant planets played a key role in shaping the habitability of the terrestrial planets.
    }
    {
    We investigate the dynamical and collisional transport processes of volatile elements by planetesimals in protoplanetary disks that host young giant planets using the HD\,163296 system as our case study. HD\,163296 is one of the best characterised protoplanetary disks that has been proposed to host at least four giant planets on wide orbits as well as a massive planetesimal disk. The goal of this study is to assess the impact of the dynamical and collisional transport on the disk and on existing and forming planetary bodies.
    }
    {
    We perform high-resolution n-body simulations of the dynamical evolution of planetesimals embedded in HD\,163296's protoplanetary disk across and after the formation of its giant planets, accounting for the uncertainty on both the disk and planetary masses as well as for the effects of aerodynamic drag of the disk gas and the gas gravity. To quantify the impact probabilities with existing and possible undiscovered planetary bodies, we process the output of the n-body simulations with well-tested statistical collisional algorithms from the study of the asteroid belt.  
    }
    {
    The formation of giant planets in the HD\,163296 system creates a large population of dynamically excited planetesimals, the majority of which originate from beyond the CO snowline. The excited planetesimals are both transported to the inner disk regions and scattered outward beyond the protoplanetary disk and into interstellar space. Existing solid planets can be enriched in volatile elements to levels comparable or larger than those of the Earth, while giant planets can be enriched to the levels of Jupiter and Saturn. 
    }
    {
    The formation of giant planets on wide orbits impacts the compositional evolution of protoplanetary disks and young planetary bodies on a global scale. The collisional enrichment of the atmospheres of giant planets can alter or mask the signatures of their formation environments, but can provide independent constraints on the disk mass. Protoplanetary disks with giant planets on wide orbits prove efficient factories of interstellar objects. 
    
    }


   \maketitle
%

\section{Introduction}

Multi-planet systems are the ones that most closely resemble our own Solar System and are the natural starting point to search for planets where habitable environments could form like on Earth. The study of the Solar System, however, has shown us how the formation and evolution of multi-planet systems is shaped in many and often subtle ways by the multifaceted interactions between their components and their surrounding environment, from the interplay between young planets and the gas and dust of their native protoplanetary disk to that between planets and minor bodies in mature planetary systems. The appearance of the giant planets in the outer Solar System, in particular, has been shown to dynamically excite the planetary bodies within their wide gravitational reach and to be responsible for multiple processes controlling the transfer of volatiles and organics from the outer to the inner Solar System and, ultimately, to Earth \citep[e.g.][]{Turrini2011,Turrini2014,Raymond2017,ronnet2018,pirani2019}. 

Giant planet-driven dynamical transport and impact contamination are processes observed being still active in the Solar System, as exemplified by the flux of Jupiter-family comets presently reaching the inner Solar System \citep[e.g.][]{Morbidelli2008,Dones2015,Galiazzo2016} and impacting Jupiter itself \citep{Taylor2004,Turrini2015,Hueso2018}. These processes were significantly more intense at the time of the solar nebula due to the larger population of small bodies existing at the time \citep{Turrini2011,Raymond2017,pirani2019}. The resulting higher impact frequencies between planetesimals allow for the systematic collisional implantation of volatiles and organics in the rocky bodies existing in the inner Solar System with efficiencies comparable to those required by the Earth's water abundance \citep{rubin2005, Turrini2014,turrini2018,Trigo2019SS}, as well as allowing for the widespread fragmentation, compaction and compositional mixing processes recorded by brecciated meteorites \citep{bischoff2006,bischoff2018,trigo-rodrigues2006}. In parallel, the higher impact frequencies between the excited planetesimals and the giant planet that cause their excitation has been shown to result in the alteration of the atmospheric compositions of the giant planets themselves (see \citealt{Turrini2015,ronnet2018,shibata2022} and \citealt{Turrini2015,Seligman2022,SainsburyMartinez2024} for the generalisation to the case of exoplanetary systems). Finally, carbonaceous planetesimals that do not undergo collisions during their dynamical transport will see their orbits circularised by the aerodynamic drag with the nebular gas and will become part of the population of minor bodies in the inner Solar System \citep{Raymond2017,OBrien2018,pirani2019} where they can be accreted by the growing terrestrial planets \citep{chambers2001,Obrien2006,alexander2017}.

While the dominant mechanism responsible for the primordial inward transport of volatiles and organics in the Solar System is still a matter of debate (e.g. whether it was controlled by Jupiter's mass growth alone or mainly sculpted by its migration, see \citealt{turrini2018,OBrien2018} and references therein), a growing body of observational evidence supports the role of dynamical transport in shaping the present-day characteristics of the Solar System. In particular, due to the volatile nature of ammonia the presence of ammoniated minerals on Ceres \citep{DeSanctis2015} and other large main belt asteroids \citep{Kurokawa2022} has been interpreted as an indication that they either originated in colder regions of the solar nebula, or accreted material transported inward from such outer orbital regions. The presence of dark comets among the near-Earth objects originating from the inner main belt \citep{Taylor2024} is also consistent with the implantation of cometary material from the outer Solar System. 

In parallel, the presence of cometary and water-rich xenoliths in carbonaceous chondrite meteorites \citep{Nittler2019,Kebukawa2020} and of apatite minerals in the eucritic meteorites originating from the basaltic surface of Vesta \citep{Sarafian2013,Sarafian2014}, one of the oldest bodies known in the Solar System \citep{Bizzarro2005,Schiller2011,Consolmagno2015}, as well as in angrite basaltic meteorites \citep{Sarafian2017} argue for this transport process having been active already during the first few million years of the life of the Solar System, after the formation of the first generation of planetesimals.

The study of these processes in the case of exoplanetary systems is made complex by the fact that many of these systems show indications that their original architectures were altered by phases of dynamical instability and chaotic evolution triggered by planet-planet scattering \citep{Laskar2017,Zinzi2017,Turrini2020,Gajdos2023}. Even in the case of young planetary systems unaltered by chaos \citep[e.g.][]{Damasso2024,Mantovan2024,Mantovan2024b}, the degeneracy in the initial conditions caused by the current lack of constraints on the disk-driven migration of their planets hinders this kind of studies. In recent years, ALMA observations have been providing an increasingly detailed view of protoplanetary disks, revealing in many of them features suggestive of the presence of forming planets at large distances from their host star \citep[e.g.][]{ALMA2015,Isella2016,fedele2017,Long2018,Andrews2018} and recently bringing to light the first direct detection of a young exoplanet embedded in its native disk \citep{Keppler2018, Muller2018}. In parallel, population studies of protoplanetary disks across different star forming regions revealed the possible resurgence of their dust abundance at ages of 2--3\,Myr \citep{testi2022}, which suggests that many of them could be host to both giant planets and massive planetesimal disks whose dynamical and collisional interactions replenish the disk dust through more intense versions of the collisional cascades that shape debris disks \citep{Bernabo2022}.

HD\,163296 is estimated to be 5--6\,Myr old \citep{Isella2016,Wichittanakom2020} and its protoplanetary disk is currently one of the best characterised ones, making it an ideal case study to investigate the dynamical transport of the astrobiologically-important volatile elements oxygen (O), carbon (C) and nitrogen (N) outside the boundaries of our Solar System. HD\,163296 has been proposed to host at least four giant planets on wide orbits \citep{Isella2016,Pinte2018,zhang2018,Teague2018,Teague2019,izquierdo2022,alarcon2022,garrido-deutelmoser2023} whose estimated masses, however, fall below the threshold for their direct detections \citep{guidi2018,mesa2019}. Furthermore, ALMA observations showed an unexpectedly high dust abundance in the disk region inward of the giant planets, where dust transport from the outer disk should be hindered by the dynamical barrier created by the planets themselves \citep{Isella2016,Liu2018,guidi2022}. The planet formation study by \cite{Turrini2019} revealed that both the presence and location of this enhanced dust region are naturally explained by the collisional erosion of planetesimals if HD\,163296's disk hosts an extended and massive population of exocomets dynamically excited by giant planets. The same study highlighted how part of the exocometary population could penetrate the innermost 10-20\,au of the protoplanetary disk, although the amount of delivered mass was not quantified, and how a large fraction of the excited planetesimals could be ejected into interstellar space \citep{Turrini2019}. The latter estimate, however, was based on a fixed ejection distance linked to the extension of the disk gas without distinguishing between planetesimals ejected into interstellar space or scattered into an extended disk similarly to what is observed in the trans-neptunian region of the Solar System \citep[e.g.][]{Morbidelli2008}.

In this work we revisit the HD\,169236 system so as to explore in more detail the process of dynamical excitation and transport of planetesimals therein, taking advantage of the updated information on the disk and stellar characteristics \citep{isella2018,BoothA2019,Kama2020,Wichittanakom2020}, on the planetary architecture \citep{Pinte2018,guidi2018,zhang2018,Teague2019,mesa2019,izquierdo2022,alarcon2022,garrido-deutelmoser2023} and on the distance of HD\,163296 from the Sun \citep{BailerJones2018,bailerjones2021,gaia2020}. Our main goal is twofold: i/ we want to assess the impact of the dynamical excitation on the planetesimals for the transport of the volatile elements C, O, and N, and its implications for the disk itself and the planets it contains (both proposed and hypothetical ones);

ii/ we aim to quantify the efficiency of this class of wide planetary systems in creating interstellar objects like the two that recently crossed the Solar System, 1I/Oumuamua and 2I/Borisov (see \citealt{MoroMartin2018, BailerJones2020, Pfalzner2021}). 

The rest of the paper is organised as follows: Section \ref{methods} reports the updated observational constraints on the HD\,163296 system and the methods we used to simulate its formation and early evolution. Section \ref{sec:results} presents the results in terms of dynamical transport and ejection of planetesimals and the collisional contamination of possible planets embedded in the disk. In Section \ref{sec:discussion} we discuss the implications of our results beyond the aspects we directly simulated, while in Section \ref{sec:conclusions} we summarise the conclusions of this work.

\section{Observational Constraints and Numerical Methods}\label{methods}

We simulate the formation of the giant planets embedded in the protoplanetary disk of HD\,163296 and their impact on the surrounding planetesimal disk with the N-body code \texttt{Mercury-Ar$\chi$es} \citep{Turrini2019,Turrini2021}. The setup of the n-body simulations builds on the one described in \cite{Turrini2019} accounting for the current constraints on the disk gas mass from \citet{BoothA2019} and \citet{Kama2020} and the inclusion of the outermost planetary companion proposed by \citep{Pinte2018}. Specifically, we consider two different disk gas masses and two sets of planetary masses for a total of four combinations of planetary and disk parameters (see Table \ref{tab:sim_param}). The stellar mass and age of HD\,163296 are always set to 1.95\,M$_\odot$ and 6\,Myr following \cite{Wichittanakom2020}. Since these stellar, disk and planetary data have been computed using the Gaia DR2 distance of HD\,163296 from \citet{BailerJones2018}, throughout this study we will keep using the DR2 distance of 101.5\,pc instead of the more recent EDR3 distance of 100.5\,pc from \citet{gaia2020} and \citet{bailerjones2021} for consistency. This choice has no impact on the results of the study as the changes in the masses and the distances within HD\,163296's system would be limited to about 1\%, therefore significantly smaller than the observational uncertainties on the relevant values.

For the disk mass we consider a ``high disk mass'' (HDM) scenario based on the gas mass estimated by \cite{BoothA2019} with ALMA using $^{13}$C$^{17}$O, where the total gas mass is 0.215\,M$_\odot$, and a ``low disk mass'' (LDM) scenario based on the gas mass estimates of \cite{Isella2016} and \cite{Kama2020} using ALMA and \textit{Herschel} observations, respectively, where the total gas mass is set to 0.05\,M$_\odot$. The protoplanetary disk is modelled adopting the gas surface density 
\begin{equation}
\Sigma(r)=\Sigma_{0} \left( \frac{r}{r_{0}} \right)^\gamma e^{\left[-\left(\frac{r}{r_{0}} \right)^{\left( 2-\gamma \right)}\right]}
\end{equation}
where $\gamma=0.8$ \citep{Isella2016}, r is the radial distance from the central object while the characteristic radius, $r_0=137.3$\,au, is obtained by scaling the original characteristic radius of $r_0=165$\,au from \cite{Isella2016} by the DR2 stellar distance from \cite{BailerJones2018}. The gas surface density $\Sigma_{0}$ is 19.4\,g\,cm$^{-2}$ in the HDM scenario and 4.5\,g\,cm$^{-2}$ in the LDM scenario. The disk gas mass is assumed to be in steady state and does not decline over time across the simulations, a reasonable approximation given that the mass of gas accreted by the giant planets amounts at most to 8\% of the disk mass (i.e. the combination of smallest disk mass and highest planetary masses) and the observed high mass-loss rates due to molecular wind have been interpreted as the onset of the disk dispersal phase \citep{Klaassen2013}. The disk temperature profile on the midplane is 
\begin{equation}
T(r)= T_{0}\,r^{-0.5}
\end{equation}
where $T_{0}=220$\,K after scaling the original temperature profile from \citet{Isella2016} by the DR2 stellar distance \citep{BailerJones2018}.

For the planets we adopt a conservative approach and focus on the three giant planets suggested to be responsible for the dust gaps between 40 and 140\,au \citep{Isella2016,Liu2018,zhang2018,Teague2018,Teague2019,izquierdo2022,alarcon2022} and the fourth outermost giant planet whose presence was proposed based on the observations of the gas dynamics \citep{Pinte2018}. We do not consider the hypothesised presence of an additional giant planet responsible for the tentative detection of a gap at about 10\,au \citep{isella2018,zhang2018}, although we will explore the implications of our results for such putative planet in Sect. \ref{sec:enrichment}, nor the proposed scenario invoking the presence of two giant planets instead of one in the dust gas at 50\,au and a possible multi-resonant orbital architecture between them and the giant planets at 86 and 137.7\,au \citep{garrido-deutelmoser2023}. For the planetary masses, we consider again a ``low planetary mass'' (LPM) scenario, where the masses of the three inner giant planets are from \cite{Liu2018} and that of the outermost giant planet from \cite{Pinte2018}, and a ``high planetary mass'' (HPM) scenario, where the masses of the three inner giant planets are from \cite{Teague2018,Teague2019} and that of the outermost giant planet from \cite{Pinte2018}. In the LPM scenario the masses originally estimated by \cite{Liu2018} based on the radial extensions of the gaps in the disk dust distribution have been rescaled to the DR2 stellar distance measured by GAIA \citep{BailerJones2018}. These scenarios for the planetary masses encompass the range of values proposed by the different studies \citep{Isella2016,Liu2018,zhang2018,Teague2018,Teague2019,izquierdo2022,alarcon2022} and are consistent with the upper limits posed by the non-detection of the planets by direct imaging \citep{guidi2018,mesa2019,huelamo2022}. The disk and planetary parameters adopted in the four scenarios are summarised in Table \ref{tab:sim_param}.

\begin{table}
	\centering
	\caption{Simulation parameters. All the values are estimated based on the DR2 distance of HD\,163296 of 101.5\,pc \citep{BailerJones2018}. The disk high-mass estimate is from \citet{BoothA2019} while the disk low-mass one is from \citet{Isella2016} and \citet{Kama2020}. The planetary low-mass scenario is based on \citet{Liu2018} and \citet{Pinte2018} while the high-mass one is based on \citet{Teague2018} and \citet{Pinte2018}.}
	\label{tab:sim_param}
	\resizebox{\columnwidth}{!}{\begin{tabular}{c|ccccc} 
		\hline
		 \textbf{Disk-Planet}  & \textbf{Disk Mass} & \textbf{Planet b} & \textbf{Planet c} & \textbf{Planet d} & \textbf{Planet e}  \\
        \textbf{scenario}      & M$_{\odot}$ & M$_{\rm J}$ & M$_{\rm J}$ & M$_{\rm J}$ & M$_{\rm J}$ \\
       \hline
        HDM-HPM & 0.215 & 0.5  & 0.83 & 1.1  & 2.0 \\
        HDM-LPM & 0.215 & 0.34 & 0.38 & 0.48 & 2.0 \\
        LDM-HPM & 0.05  & 0.5  & 0.83 & 1.1  & 2.0 \\
        LDM-LPM & 0.05  & 0.34 & 0.38 & 0.48 & 2.0 \\
        \hline
        \multicolumn{2}{l}{\textbf{Semimajor axis (au)}} & 50 & 83 & 137.3 & 260\\
        \hline		
		
	\end{tabular}}
\end{table}

The n-body simulations model the dynamical evolution of a disk of planetesimals embedded within HD\,163296's gaseous disk under the effects of the mass growth of the four forming giant planets, of the aerodynamic drag by the disk gas and of the disk gravity. Following \cite{Turrini2019,Turrini2021}, planetesimals are modelled as test particles possessing inertial mass and no gravitational mass, meaning that the test particles are affected by the disk gas but do not influence each other nor the giant planets. 

The inertial mass is computed assuming a characteristic diameter of 100\,km \citep[see][]{KlahrSchreiber2016,johansen2017} and bulk density of 1\,g/cm$^3$ \citep[see][]{Turrini2019,Turrini2021}. The damping effects of aerodynamic gas drag on the planetesimals are simulated following the treatment from \citet{brasser2007} with updated drag coefficients from \citet{nagasawa2019} accounting for both the Mach and Reynolds numbers of the planetesimals. The exciting effects of the disk gravity are simulated based on the analytical treatment for axisymmetric disks by \citet{ward1981} following \citet{marzari2018} and \citet{nagasawa2019}. We refer interest readers to \cite{Turrini2021} for additional details.

The formation of the giant planets is modelled over two growth phases using the parametric approach from \citet{Turrini2011,Turrini2019}. The first phase accounts for their core growth and subsequent capture of an expanded atmosphere \citep[e.g.][]{Bitsch2015,johansen2019,DAngelo2021}. The planetary mass evolves as 
\begin{equation}
M_{p}(t)=M_{0}+\left( \frac{e}{e-1}\right)\left(M_{1}-M_{0}\right)\left( 1-e^{-t/\tau_{p}} \right)
\end{equation} 
where M$_{0}=0.01$\,M$_{\oplus}$ is the initial mass of the core, M$_{1}$ is the final cumulative mass of the core and its expanded atmosphere at the end of the first growth phase (see \citealt{Turrini2021} for further discussion), $e$ is the Euler number, $t$ is the time, and $\tau_{p}$ is the duration of the first growth phase. The second phase of mass growth accounts for the runaway gas accretion of the two giant planets, where their mass evolves as 
\begin{equation}
M_{p}(t)=M_{1}+\left( M_{2} - M_{1}\right)\left( 1-e^{-(t-\tau_{p})/\tau_{g}}\right)
\end{equation} 
with M$_{2}$ being the final mass of the giant planets and $\tau_{g}$ the e-folding time of the runaway gas accretion process.

\begin{figure*}[ht]
	  \includegraphics[width=\textwidth]{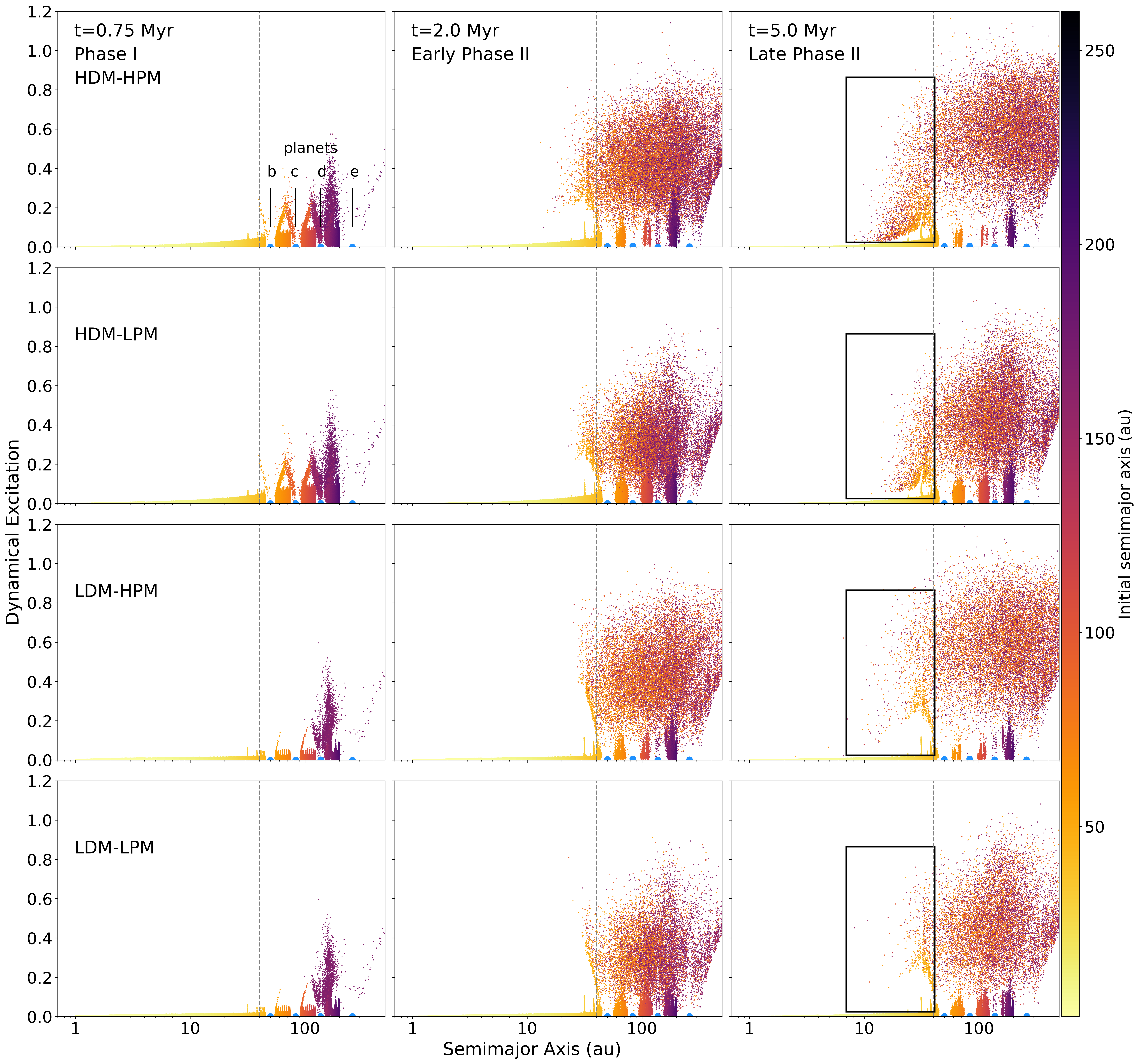} 
    \caption{Dynamical excitation of the planetesimal disk at 0.75, 2 and 5\,Myr in the four simulated scenarios. Dynamical excitation accounts for the contributions of both eccentricity and inclination and is defined as $\sqrt{e^2 + \sin^2(i)}$ \citep{Petit2001}. The colour bar indicates the region in the disk from which the planetesimals originated (in units of au). The dashed grey line at 40\,au indicate the division between inner and outer disk. The rectangles in the rightmost panels mark the ``fountain'' feature we refer to in the main text, i.e. the damping of the orbital eccentricities of the excited planetesimals by aerodynamic drag that creates the implanted planetesimals. The light blue filled circles indicate the location of the four simulated planets in the disk. \textbf{First row:} High Disk Mass - High Planetary Mass; \textbf{Second row:} High Disk Mass - Low Planetary Mass; \textbf{Third row:} Low Disk - High Planetary Mass; \textbf{Fourth row:} Low Disk Mass - Low Planetary Mass}
    \label{fig:mosaic}
\end{figure*}

As in \cite{Turrini2019}, we model the formation of HD\,163296's four giant planets as occurring in situ, i.e. without them undergoing orbital migration while growing in mass. This simplifying assumption is motivated by the fact that dynamical excitation of the planetesimal disk is dominated by the rapid mass growth and migration of the giant planets during their runaway gas accretion phase \citep{Turrini2011,Turrini2012,Turrini2019,Raymond2017,pirani2019} and recent studies indicate that the migration of giant planets during such phase is limited \citep{Tanaka2020,Paardekooper2023}. The inclusion of the migration of the planetary core  result in slightly earlier onsets of the dynamical excitation \citep{pirani2019}. 

The adopted values of M$_{2}$ for the four planets are those of their estimated masses in the HPM and LPM scenarios. For the three innermost giant planets we adopt values of $M_{1}=30$\,M$_\oplus$ \citep{Lissauer2009,DAngelo2021}, $\tau_{p}=1$\,Myr \citep{mulders2021,Bernabo2022,lichtenberg2023}, and $\tau_{g}=0.1$\,Myr \citep{Lissauer2009,DAngelo2021} to simulate their growth by core accretion. The timescale of core formation $\tau_{p}$ has been chosen to fit the temporal interval in the disk lifetime that is most favourable to the formation of giant planets \citep{savvidou2023}. For the outermost planet, we set $M_{1}=0.3$\,M$_\oplus$, $\tau_{p}=0.01$\,Myr, and $\tau_{g}=0.01$\,Myr to simulate its faster formation by disk instability \citep{DAngelo2010,Helled2014}.

During their gas accretion phase all four giant planets form gaps in the disk gas whose widths are modelled as $W_{gap} = C\cdot R_{H}$ \citep{Isella2016,marzari2018}, where the numerical proportionality factor $C=4$ is from \citet{Isella2016} and \citet{marzari2018} and $R_{H}$ is the relevant planetary Hill's radius. The gas density $\Sigma_{gap}(r)$ inside the gap evolves over time with respect to the local unperturbed gas density $\Sigma(r)$ as \begin{equation}
\Sigma_{gap}(r) = \Sigma(r)\cdot e^{\left[-\left(t-\tau_{p}\right)/\tau_{g}\right]}
\end{equation}
\citep{Turrini2021}. When planetesimals cross the gaps the effects of the gas on their orbital evolution are computed using the local gas density $\Sigma_{gap}$.

We set the spatial density of planetesimals in the N-body simulations to 1000\,particles/au, with the inner edge of the planetesimal disk at 1\,au and the outer edge at 200\,au (see \citealt{Turrini2021} for the discussion of the choice of the disk inner edge). The orbital regions corresponding to the feeding zones of the giant planets (see \citealt{DAngelo2010} and references therein) are not populated by planetesimals to account for the local solids accreted in their cores or trapped within their collapsing gas. The resulting planetesimal disks are populated by about 10$^{5}$ test particles. In order to accurately reproduce the orbital evolution of planetesimals close to the inner edge of the disk, the timestep is set smaller than 5\% of the orbital period at 1\,au \citep[e.g.][]{rein2015} by adopting a value of 15 days. Planetesimals are considered ejected from the protoplanetary disk into interstellar space once their orbits become unbound, i.e. when their eccentricity becomes equal or greater than one. We record snapshots of the evolving architecture of the planetary system every 10$^4$ years, keeping trace of the original formation region of the planetesimals. The recorded eccentricity $e$ and inclination $i$ values allow to quantify the dynamical excitation of the planetesimals as $\sqrt{e^2 +\sin{^2}(i)}$ \citep{Petit2001}.

To quantify how the solid mass is redistributed across the protoplanetary disk by the appearance of the four giant planets, we treated each test particle in the N-body simulations as a swarm of real planetesimals. Since constraining the original mass of gas and solids of HD\,163296 is a challenging task \citep{Turrini2019,BoothA2019,mulders2021}, particularly in light of the high mass loss \citep{Klaassen2013} and gas accretion rate \citep{Wichittanakom2020} presently experienced by this protoplanetary disk, we adopt the following simplified approach. The cumulative mass of each swarm is computed integrating the adopted disk gas density profile over a ring 0.1\,au wide centred on the initial orbit of the impacting particle, and multiplying the resulting gas mass by the local solid-to-gas ratio. The solid-to-gas ratio, in turn, is a function of the disk metallicity and local disk midplane temperature and is described by the simplified radial condensation profile from \citet{Turrini2023}. 

The disk metallicity is set to 1.4\% \citep{asplund2009} under the assumption of solar metallicity for consistency with the characterisation of  HD\,163296's stellar parameters by \cite{Wichittanakom2020}.
The solid-to-gas ratio is 0.5 times the disk metallicity for planetesimals formed at temperatures between 1200\,K and 140\,K, i.e. between the condensation of silicates and that of water. The solid-to-gas ratio grows to 0.75 times the disk metallicity for planetesimals formed at temperatures between 140\,K and 30\,K, i.e. between the snowlines of water and carbon monoxide, and reaches 0.9 times the disk metallicity for planetesimals formed at temperatures below 30\,K, i.e. beyond the carbon monoxide snowline.

The impacts on the proposed four giant planets forming in HD\,163296's disk rings are recorded directly during the n-body simulations. Then the formation region of the impacting particles is used to quantify the mass of the swarm of planetesimals they represent based on the methods described above. To investigate the collisional implications of the dynamical transport of planetesimals to the inner regions of the protoplanetary disk, we also compute the impact probabilities and velocities among the particles that originate within the inner unperturbed planetesimal disk and those originating from the outer excited planetesimal disk using the well-tested statistical collisional methods developed for the study of the asteroid belt \citep{Wetherill1967,Farinella1992,OBrien2011}. The transition between the two planetesimal populations is set at 40\,au based on the results of the n-body simulations (see Sects. \ref{sec:implantation} and \ref{sec:injection}). The resulting values are used to estimate the accretional fluxes of impactors on possible undiscovered inner planets hidden within HD\,163296's gas as discussed in Sect. \ref{sec:enrichment}.

\section{Results}\label{sec:results}

In all simulated systems (Figure \ref{fig:mosaic}) we can distinguish two planetesimal populations and two temporal phases concerning their dynamical evolution. One population is that of the dynamically excited planetesimals that formed in the disk region where the giant planets reside. The other population is that of the unperturbed (or limitedly excited) planetesimals that formed inward of the giant planets. As shown by Figure \ref{fig:mosaic} and introduced in Sect. \ref{methods}, the transition between the two populations occurs at about 40\,au. In the following we will refer to the first 40\,au as \textit{inner disk} and to the region beyond 40\,au as \textit{outer disk}.

The two populations remain well separated during the first temporal phase that extends until all four giant planets reach their current masses, i.e. before 1\,Myr in our simulations. In this phase the most dynamically excited region is the one between the third and fourth planets, mainly due to the assumed rapid formation of the outermost planet by disk instability (see Section \ref{methods}). During this phase the excited planetesimals mostly cross the orbital region of the giant planets and only limitedly penetrate within the inner disk. In this phase we can see a difference in the level of global dynamical excitation between high mass and low mass disks. Specifically in the HDM scenarios the stronger effect of the disk gravity adds to the planetary perturbations producing a higher level of dynamical excitation with respect to the LDM cases (see the planetesimals in the rings between the planets at t=0.75\,Myr in Fig. \ref{fig:mosaic}).

The second temporal phase starts once the giant planets complete their growth. Planetesimals in the outer disk reach very high levels of dynamical excitation (producing eccentricity values above 0.4, see also Fig. \ref{fig:inj}) and are able to systematically penetrate the innermost 40\,au where they start being significantly affected by the gas drag (see Sect. \ref{sec:injection}). During this second phase, therefore, we observe the systematic mixing between the two planetesimal populations within the inner disk (see Sect. \ref{sec:implantation}). In the LDM scenarios the effect of gas drag remains limited and only the most excited planetesimals, moving at high relative velocity with respect to the gas, see their excitation damped and have their orbits circularised. In the HDM scenarios, instead, planetesimals that are injected inside the orbit of the second planet (i.e. inwards of 80\,au) are affected by a stronger gas drag and see their eccentricity significantly damped over time (the ``fountain'' feature inward of 40 au at 5\,Myr identified by the box in Fig. \ref{fig:mosaic}). By 5\,Myr a fraction of these damped planetesimals has been stably implanted within the inner disk, in some cases reaching inwards of 10\,au (see Figs. \ref{fig:mosaic} and \ref{fig:inj}).

In the rest of this work we will identify as \textit{implanted} in the inner disk those excited planetesimals that formed in the outer disk, have pericentres in the inner disk and whose orbits underwent circularisation due to the gas drag to the point that their apocentres became less than 40\,au (see Fig. \ref{fig:inj}), i.e. the planetesimals became decoupled from the outer disk. This process causes the implanted planetesimals to mix with the original unperturbed population of the inner disk region (see Sect. \ref{sec:implantation} for details). We will use, instead, the term \textit{injected} to identify those excited planetesimals that formed in the outer disk, have pericentres in the inner disk and remain on eccentric orbits with apocentres in the outer disk (see Fig. \ref{fig:inj}), i.e. for which gas drag is not able to circularise the orbits. These planetesimals can contribute to the compositional evolution of the inner disk through collisional processes (see Sect. \ref{sec:injection}).  

The dynamical excitation process scatters a marked population of planetesimals in an extended disk beyond the orbit of planet e and causes the \textit{ejection into interstellar space} of one tenth to one fourth of the initial planetesimal disk. The three inner giant planets accrete, in total, between 4-5\% of the planetesimals initially embedded in the disk in the HDM scenarios, while in the LDM scenarios their accretion efficiency drops to 1-2\% (see Sect. \ref{sec:impacts}). The outermost giant planet, due to its faster formation and larger final mass, quickly becomes more efficient in scattering planetesimals rather than accreting them and undergoes no impacts during the simulations. 

\subsection{Implantation in the inner disk}\label{sec:implantation}

\begin{figure*}
	\includegraphics[width=\textwidth]{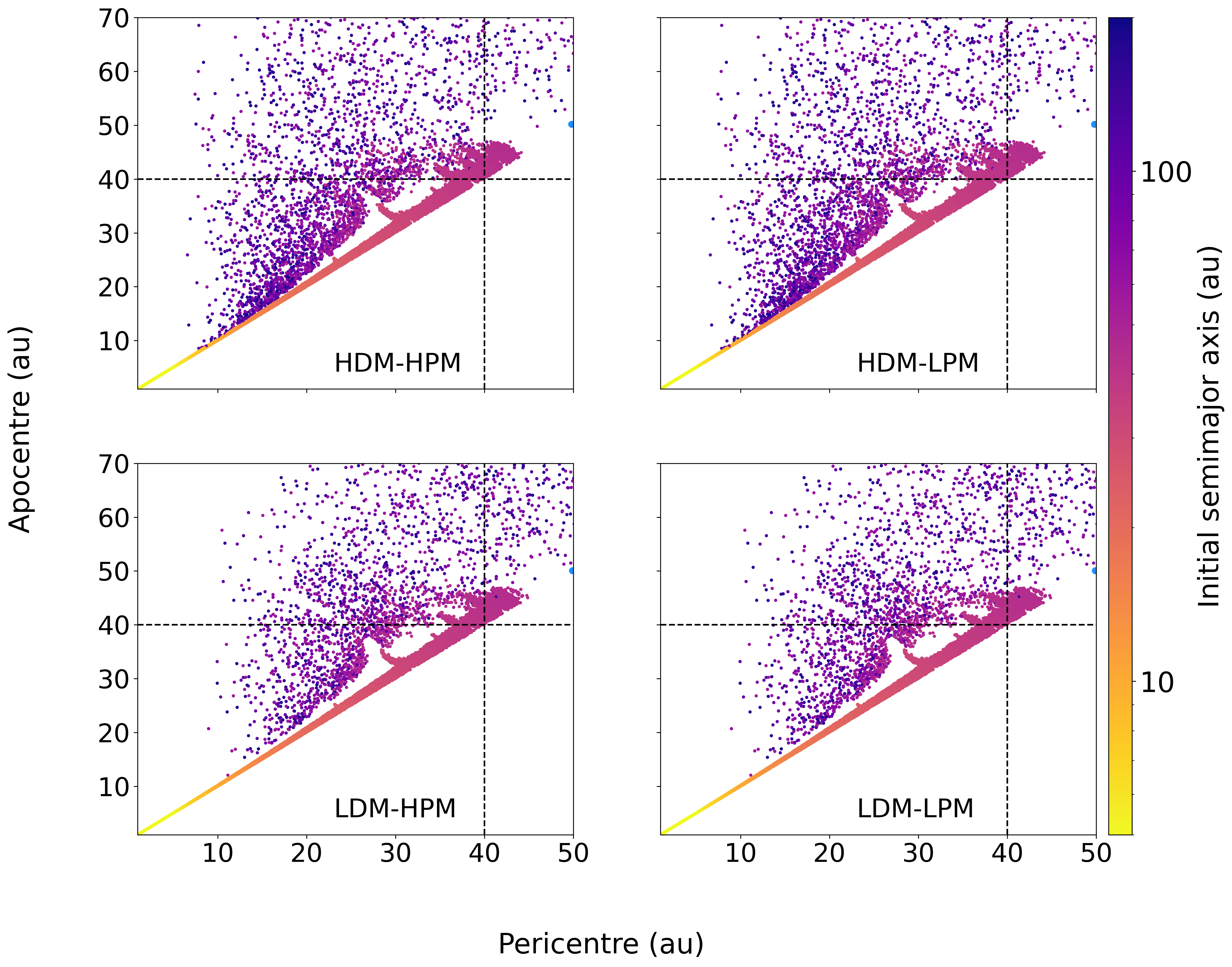}
    \caption{Pericentre vs. apocentre plot of the planetesimal disk at the end of our simulations at 5\,Myr, where we can clearly distinguish the excited and unperturbed planetesimal populations. Excited planetesimals from the outer disk can penetrate the inner disk down to a few au from the host star. Planetesimals are colour-coded based on their formation region, the colour scale is truncated at 100\,au to better resolve the inner disk. The vertical and horizontal dashed lines at 40\,au split the disk in the inner and outer region}\label{fig:inj}
   
\end{figure*}

Once the system enters its second temporal phase of dynamical excitation, the orbital evolution of the planetesimals injected in the inner disk becomes dominated by gas drag and their architecture is shaped by its damping effects, as exemplified by the presence of the ``fountain'' features that can be seen in Figure \ref{fig:mosaic}. The orbital circularisation caused by this damping process permanently implants the excited planetesimals into the inner disk and mixes them with the local unperturbed planetesimal population, similarly to what has been suggested for the asteroid belt in the Solar System \citep{Turrini2014,DeSanctis2015,Raymond2017,pirani2019}.

To assess the efficiency and the potential impact of the inward transport and implantation of planetesimals and astrobiologically-important elements, we quantify the ratio between local planetesimals and implanted planetesimals in each 10\,au-wide ring within the inner disk region (see Table \ref{tab:injected}). The planetesimals are assumed to be implanted in the rings within whose boundaries their pericentres are located at the end of the simulations. As shown in Figure \ref{fig:mosaic},
a number of implanted planetesimals are still characterised by marked eccentricities ($e>0.1$), meaning that they can spend part of their orbits outside the boundaries of the ring. Given that HD\,163296 still hosts its protoplanetary disk after the timespan covered by our simulations, our approach is equivalent to assuming that the gas drag over the subsequent life of the disk will circularise the orbits of the implanted planetesimals at their pericentres. 

\begin{figure*}
	\includegraphics[width=\textwidth]{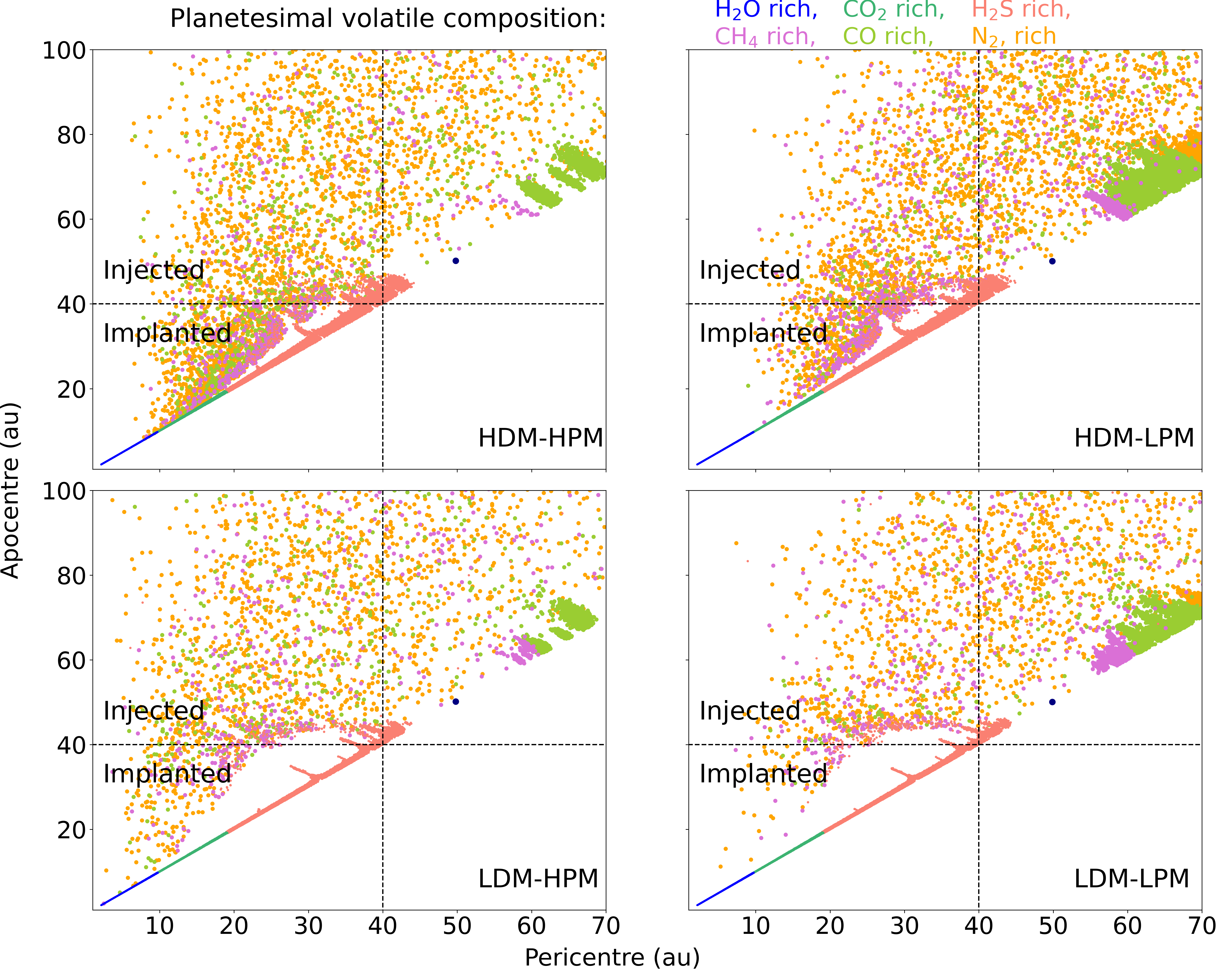}
    \caption{Same as Fig. \ref{fig:inj} with the excited planetesimals being colour--coded according to the compositional region they originate from. This compositional gradient is based on HD\,163296's temperature profile and on the condensation temperatures as ices of the main volatile molecules.}
    
    \label{fig:chem}
\end{figure*}

As shown in Figure \ref{fig:inj} and Table \ref{tab:injected}, implantation is most effective in the region between 10-30\,au, spanning two 10\,au wide rings, in the case of the HDM scenarios. In the LDM scenarios the peak efficiency is reached between 10 and 20 au. The majority of implanted planetesimals originate beyond the CO line at 60\,au with a small fraction originating instead in the region between 40 and 60\,au, i.e. between the CH$_4$ and CO snowlines (see Fig. \ref{fig:chem}). About 50--60\% of the planetesimals originating beyond the CO snowline actually come from beyond the N$_2$ snowline, enriching the local ring composition in nitrogen alongside carbon and oxygen (see Fig. \ref{fig:chem}). 

\begin{table}
	\centering
	\caption{Mass fraction of planetesimals implanted and injected into the inner disk. The mass fractions are expressed with respect to the local planetesimal ring mass in the four disk-planets scenarios.}
	\label{tab:injected}
	\begin{tabular}{c|ccccc} 
		\hline
		 & \textbf{Ring} & \multicolumn{2}{c}{\textbf{Implanted bodies}} & \multicolumn{2}{c}{\textbf{Injected bodies}}\\
      & \textbf{(au)} & \textbf{HPM} & \textbf{LPM} & \textbf{HPM} & \textbf{LPM}\\
        \hline
        \parbox[t]{2mm}{\multirow{4}{*}{\rotatebox[origin=c]{90}{HDM}}}
       & 40 -- 30 & <1\% & <1\% & 46\% & 37\%\\
       & 30 -- 20 &  10\% &  8\% & 17\% & 11\%\\
       & 20 -- 10 & 11\% &  3\% &  5\% &  2\%\\
       & 10 -- 0.1 & <1\% & <1\% & <1\% & <1\%\\
        \hline
        \parbox[t]{2mm}{\multirow{4}{*}{\rotatebox[origin=c]{90}{LDM}}}
       & 40 -- 30 &  0\% &  0\% & 38\% & 25\%\\
       & 30 -- 20 & <1\% & <1\% & 19\% &  10\%\\
       & 20 -- 10 &  3\% &  1\% &  7\% &  2\%\\
       & 10 -- 0.1 & 1.3\% & <1\% & 1\% & <1\%\\

        \hline
	\end{tabular}
\end{table}

\begin{figure}
	\includegraphics[width=\columnwidth]{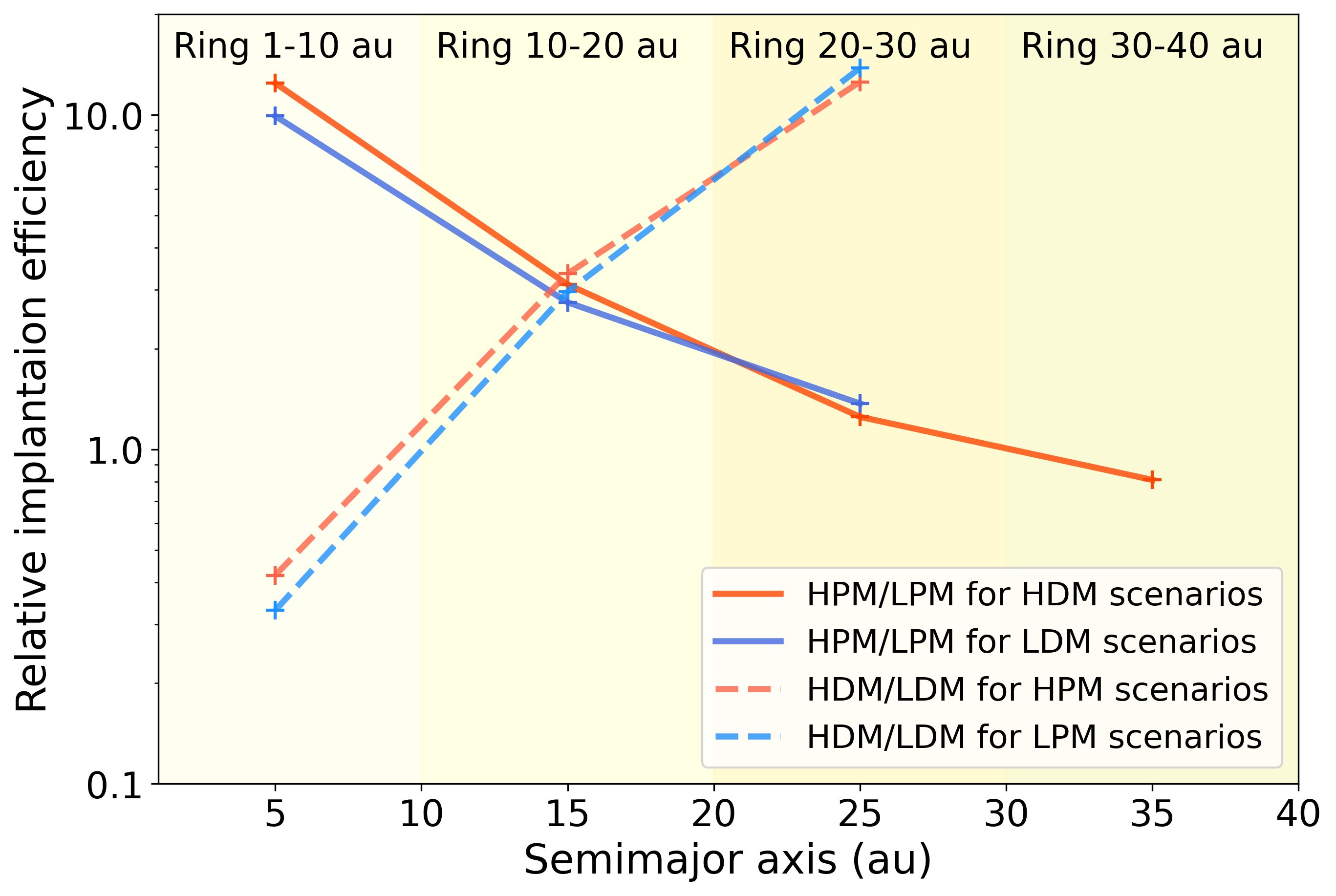}
    \caption{Relative efficiency between the HPM and LPM scenarios (solid curves) and between the HDM and LDM scenarios (dashed curves) in implanting planetesimals in the inner disk. The orange color identifies the high-mass disk when comparing the efficiencies of the HPM and LPM scenarios, and the high-mass planets when comparing the efficiencies of the HDM and LDM scenarios. The blue curves identify the LDM and LPM scenarios, respectively. High relative efficiencies in the curves can still be associated to limited absolute efficiencies in Table \ref{tab:injected}. Increasing the disk mass enhances the implantation efficiency in the part of the inner disk closer to the giant planets, while increasing the planetary masses enhances the implantation efficiency closer to the host star.} \label{fig:variations}
\end{figure}

In the HDM-HPM scenario the giant planets can implant planetesimals between 10-20\,au and 20-30\,au with similar efficiencies (Table \ref{tab:injected}), while their less massive counterparts in the HDM-LPM scenario are three times less efficient in implanting planetesimals between 10 and 20\,au than between 20 and 30\,au  (Table \ref{tab:injected}). The implantation efficiency in the outermost (30--40\,au) and innermost (0--10\,au) rings proves always negligible in the HDM scenarios. In the LDM scenarios the most enriched ring is that between 10 and 20\,au, which sees its mass increased by 3\% and 1\% in the HPM and LPM scenarios, respectively. Differently from the HDM case, however, in the LDM-HPM scenario also the innermost ring (0--10\,au) sees its mass increased by 1.3\% thanks to the interplay between the stronger planetary perturbations and the weaker gas drag that allows planetesimals to penetrate deeper in the inner disk. The comparison of the implantation efficiencies in the different orbital regions as a function of the disk and planetary masses is shown in Fig. \ref{fig:variations}. Specifically, the solid lines in Fig. \ref{fig:variations} show the ratios between the implantation efficiencies for the two different sets of planetary masses while keeping the disk mass constant, while the dashed lines show the ratios for the two different disk masses while keeping the planetary masses constant. As can be immediately seen from the slopes of the curves in Fig. \ref{fig:variations}, increasing the disk mass enhances the implantation efficiency in the part of the inner disk closer to the giant planets, while increasing the planetary masses enhances the implantation efficiency closer to the host star.

The enrichment in mass due to implantation, while moderate even in the most efficiently implanted ring, is comparable to that estimated to have occurred in the inner Solar System and the asteroid belt following the formation of Jupiter and Saturn \citep[1-10\%][]{Raymond2017,ronnet2018} while affecting an orbital region an order of magnitude larger. In other words, HD\,163296's protoplanetary disk is characterised by higher degrees of dynamical excitation and compositional remixing than the Solar System. The higher abundance of N in the implanted planetesimals \citep{Oberg2021} makes the enrichment in this element much more significant than what the implanted mass suggests. 

More generally, due to the important roles of N$_2$ and CO as carriers of nitrogen and carbon \citep{Oberg2021}, the enrichment in these elements of the inner disk region (0--40\,au) proves always larger than the one in H$_2$O and volatile elements estimated as having occurred in the asteroid belt and the inner Solar System due to cometary impactors \citep[0.1--0.01\%, see][and references therein]{Turrini2014,Sarafian2017,turrini2018}. Finally, the mass implanted between 10 and 30\,au in the HDM scenario and between 0 and 20\,au in the LDM scenario is comparable in magnitude or greater than that estimated for the \textit{late veneer} that affected the planetary bodies in the asteroid belt and the inner Solar System \citep[see][and references therein]{Day2012,Day2016,turrini2018}. 

\subsection{Injection in the inner disk}\label{sec:injection}

Alongside the effect of dynamical implantation we also evaluated the efficiency of dynamical injection and collisional delivery of planetesimals to the inner regions of the disk, similarly to the case of the cometary flux in the inner solar system \citep{Turrini2014,turrini2018}. Since the dynamical injection process depends only on the orbital eccentricity of the excited planetesimals, it involves significantly more bodies than the implantation process and is particularly efficient with the highest planetary masses considered here (see Table \ref{tab:injected}).

While lower disk masses result in decreased damping efficiency of gas drag and should allow for the injection of a larger population of high eccentricity planetesimals, in Table \ref{tab:injected} we observe this behaviour only with the HPM cases. For the LPM cases the HDM scenario is always characterised by higher injection efficiencies than the respective LDM counterpart due to the effects of the stronger disk gravity. The exciting effects of the local disk gas increase the number of planetesimals that cross the orbital region of the giant planets and get their eccentricity increased during close encounters (see Fig. \ref{fig:mosaic} at 0.75\,Myr).

The injection efficiency peaks in the outermost ring of the inner disk, i.e. between 30 and 40\,au, and decreases monotonically when moving to rings closer to the star. The injection efficiency decreases a factor of two faster for the LPM scenarios than for the HPM scenarios. Specifically, in the LPM scenarios it drops by a factor of 10-15 moving from the ring between 30-40\,au to the ring between 10-20\,au, while over the same orbital range the efficiency of the HPM scenarios drops only by a factor of 4-6. The orbital region between 20 and 40\,au is crossed by a mass flux of planetesimals representing 10-30\% of the local mass of planetesimals. 

As shown in Fig. \ref{fig:inj}, excited planetesimals can reach the disk region inward of or very close to 10\,au in all scenarios. In the HDM-HPM scenario the mass of injected planetesimals reaches up to 1\% of the local planetesimal population, while in all other scenarios the injected mass is always less than this value (see Table \ref{tab:injected}). Notwithstanding this, the efficiency of the injection process in HD\,163296 proves comparable to that characterising the scattering process in the inner Solar System triggered by the formation of Jupiter and Saturn \citep{Raymond2017,ronnet2018} while affecting an orbital region an order of magnitude wider.

\subsection{Collisional enrichment and contamination of possible inner planets}\label{sec:enrichment}
\begin{figure*}
	\includegraphics[width=\textwidth]{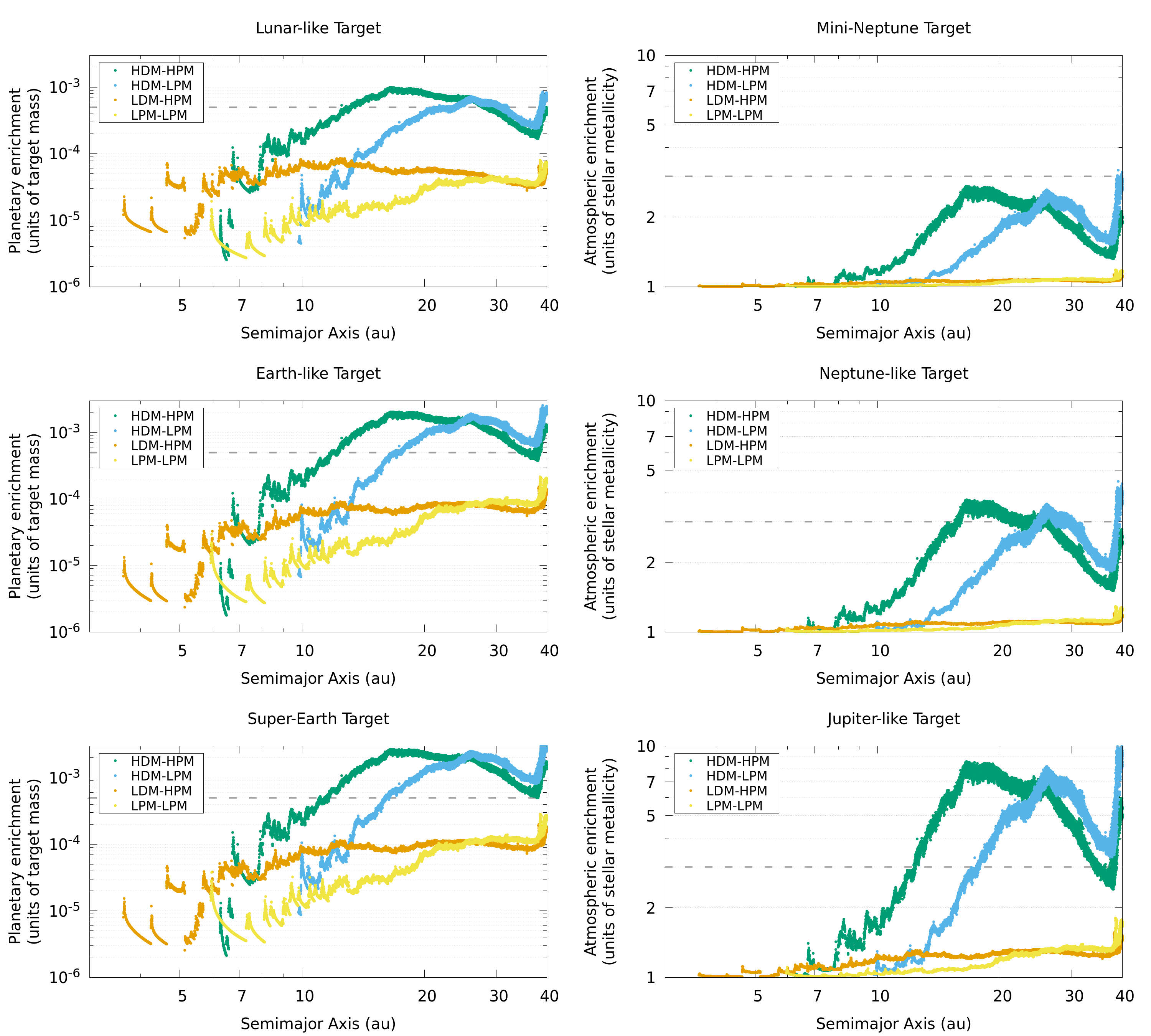}
    \caption{Enrichment of possible undiscovered planets in the inner disk by the impacts of planetesimals that formed in the outer disk for all four simulation scenarios (with blue we denote the HDM-HPM scenario, with green the HDM-LPM one, with orange the LDM-HPM one and with yellow the LDM-LPM one). \textbf{Left column:} from top to bottom, normalised mass fraction of accreted material onto Lunar-like, Earth-like and super-Earth targets, with the horizontal dashed line marking the water mass fraction on Earth (5$\times$10$^{-4}$, \citealt{Morbidelli2000}). \textbf{Right column:} from top to bottom, atmospheric enrichment of mini-Neptune, Neptune-like an Jupiter-like targets assuming a solar-like metallicity atmosphere, with the horizontal dashed line marking Jupiter's 3x solar atmospheric metallicity \citep{atreya2018}.}\label{fig:seedImp}
\end{figure*}

As introduced in Sect. \ref{methods}, from the analysis of the observations from the DSHARP ALMA survey \citep{Andrews2018} \citet{zhang2018} and \citet{isella2018} report evidence for an additional gap at 10\,au within HD\,163296's protoplanetary disk, not seen in the previous works on the system (e.g. \citealt{Isella2016}). In particular, \citet{zhang2018} postulate that such a ring can be the result of a new planet with mass ranging between 0.35--0.7\,M$_J$ embedded in the disk. While we do not account for this possible planet in our simulations, in light of the estimated efficiency of the injection process this result prompted us to look into the collisional enrichment and contamination effects of the planetesimal flux crossing the inner disk on possible planetary bodies existing within 40\,au, following \cite{Turrini2014} and \cite{Turrini2015,turrini2018}. Specifically, following \citet{Turrini2015} we aim both to quantify the increase in the abundance of the volatile elements (enrichment) and to assess whether this enrichment can significantly alter or mask the compositional signatures of the formation process (contamination). In order to calculate the numbers of potential impacts on these assumed targets, we focused on the orbital architectures of the planetesimal disks in the final snapshot of our simulations and computed the intrinsic impact probability \citep[see Sect. \ref{methods} and][]{Wetherill1967,Farinella1992} between all possible target-impactor pairs. 

We considered as targets the orbits and positions of all planetesimals originating within 40\,au as a proxy for planetary sized objects, while we considered as impactors all the planetesimals originating beyond 40\,au whose pericentres reach inward of 40\,au. We focused on six impact scenarios based on templates from both the Solar System\footnote{ \url{https://nssdc.gsfc.nasa.gov/planetary/factsheet/}} and exoplanets\footnote{\url{https://exoplanetarchive.ipac.caltech.edu/}} spanning the mass range from small planetary embryos to fully-formed massive planets. Specifically, we assumed targets with Lunar-like (0.01\,M$_{\oplus}$), Earth-like (1\,M$_{\oplus}$), Neptune-like (17\,M$_{\oplus}$) and Jupiter-like (318\,M$_{\oplus}$) masses as well as targets modelled after the super-Earth GJ\,486b (M$_{t}$=3\,M$_{\oplus}$, R$_{t}$=1.34\,R$_{\oplus}$, v$_{esc}$=16.7 km/s) and the mini-Neptune TOI-1260b (M$_{t}$=8.56\,M$_{\oplus}$, R$_{t}$=2.41\,R$_{\oplus}$, v$_{esc}$=21.1 km/s), based on the classification from \citet{fulton2017}. We focused our evaluation on a temporal window of 1\,Myr, i.e. the difference between the timespan covered by our simulations and the estimated age of HD\,163296 \citep{Wichittanakom2020}, to account for the possible slower or later formation of the putative inner planets.

To compute the impact fluxes from the intrinsic impact probabilities we used the collisional formula  \citep[see][and references therein]{OBrien2011}:
\begin{equation}
    N_i= P_i \times (R_{t,eff} + R_{pl})^2 \times \Delta t \times N_{pl}
\end{equation}    
where N$_i$ is the number of impacts on the target body, P$_i$ is the intrinsic impact probability (see Sect. \ref{methods} and \citealt{Wetherill1967,Farinella1992}, the value includes the $\pi$ term of the cross section), R$_{pl}$ is the projectile radius in km (here fixed at 50\,km), $\Delta$t the temporal interval considered (here 1\,Myr), N$_{pl}$ is the number of planetesimals that are injected into the inner disk, and R$_{t,eff}$ is the target's effective radius in km defined as
\begin{equation}
    R^2_{t,eff}= R^2_t \times \left(1+ \left(\frac{\nu_{esc}}{\nu_{imp}}\right)^2\right) 
\end{equation}
where $R_t$ is the target's physical radius, $\nu_{esc}$ and $\nu_{imp}$ are the escape and impact velocity respectively. The term $1+\left(\frac{\nu_{esc}}{\nu_{imp}}\right)^2$ is the gravitational focusing factor \citep{Safronov1972}. 

The resulting numbers of impacts were multiplied by the inertial mass of the colliding planetesimals and normalised depending on the mass of the target body. In the case of the Mini-Neptune-like, Neptune-like and Jupiter-like targets, the mass flux of impactors is normalised to the mass of heavy elements contained in an atmospheric shell of solar metallicity accounting for 20\% of 1.5 the target mass following \citet{Turrini2015}. The heavy elements in these atmospheric shells amount to about 0.02\,M$_\oplus$ for the Mini-Neptune targets, 0.05\,M$_\oplus$ for the Neptune-like targets and 0.9\,M$_\oplus$ for the Jupiter-like targets. These values are used to compute the amount of heavy elements required to reach an atmospheric enrichment 3x solar as Jupiter \citep{atreya2018}, to which we compare the enrichments produced by collisional contamination. 

For the Lunar-sized, Earth-sized and Super-Earth-sized targets, the cumulative mass flux of planetesimals is instead normalised to the whole target mass. As reference value to compare the magnitude of the resulting late accretion with Solar System analogues, we adopt the water mass fraction estimated for the Earth ($5\times10^{-4}$ planetary masses, \citealt{Morbidelli2000}). We need to emphasise, however, that Earth's water enrichment does not translate into an unequivocal late accretion mass value due to the uncertainty on the source of Earth's water. Specifically, Earth's water enrichment requires the accretion of a mass of planetesimals twice as large if these planetesimals are cometary in nature and have half their mass as water ice \citep[e.g.][and references therein]{Turrini2018b}. If the source of Earth's water was instead linked to carbonaceous asteroids, whose water content amounts to about 10\% in mass  \citep[e.g.][and references therein]{Turrini2018b}, the mass of planetesimals required to produce Earth's water abundance would be ten times as large. Because of this uncertainty and the lack of detailed information on the chemical setup of HD\,163296's disk, Earth's water content is adopted as a convenient reference but the outcomes of the comparison should not be over-interpreted. 

The effects of the collisional contamination by impacting planetesimals are summarised in Fig. \ref{fig:seedImp}\footnote{The bulk planetary enrichments in Fig. \ref{fig:seedImp} can be converted to crustal enrichments by multiplying the values in the plots by a factor of 20. This conversion factor is equivalent to normalizing the late accretion of planetesimals by a crustal mass amounting to $1\%$ the planetary mass, by analogy with the case of the Earth adopting an average crustal thickness of 40 km and an average crustal density of 2800 kg/m$^3$ \citep{Christensen1995}. A bulk enrichment of $5\times10^{-4}$ is therefore equivalent to a crustal enrichment of $1\%$, i.e. the magnitude of the late veneer experienced by asteroids in the Solar System \citep{Turrini2018b}}. In the HDM scenarios Lunar-like, Earth-like and super-Earth targets can become enriched in ices from the outer disk to levels comparable to Earth's water enrichment, while in the LDM scenarios the enrichment of this class of targets is always smaller than this reference value by at least a factor of a few (left panels in Fig. \ref{fig:seedImp}). The highest enrichments are reached by target planets beyond 10-15 au in the HPM scenarios and beyond 15-20 au in the LPM scenarios, with the magnitude of the enrichment increasing proportionally to the gravitational cross-section of the target planet (e.g. Super-Earths experience higher enrichments than Earth-like and Lunar-like targets). In these orbital regions, the amount of outer disk planetesimals accreted by these target planets falls between the cases of cometary and carbonaceous asteroidal impactors as the source of Earth's water discussed above \citep{Turrini2018b}.

Also in the case of the three more massive targets (right panels in Fig. \ref{fig:seedImp}), the larger gravitational cross-section of the Jupiter-like targets allows for a greater enrichment than their Neptune-like and mini-Neptune counterparts. Specifically, in the HDM scenarios the atmospheric enrichment of the Jupiter-like targets is always super-solar beyond 10 au and can grow to about 9x (i.e. Saturn's atmospheric enrichment, \citealt{atreya2018}) between about 15 and 30\,au as well as around 40 au. The atmospheres of Neptune-like planets can become 3x to 4x enriched in heavy elements with respect to the solar composition (i.e. as enriched as Jupiter) between 10 and 30\,au in the HPM cases and between 20 and 30\,au in the LPM cases (see Fig. \ref{fig:seedImp}), as well as around 40 au in both cases. The atmospheres of mini-Neptune targets are the least enriched yet they range between 2x and 3x solar beyond 15 au in the HPM case and 20 au in the LPM case. In the LDM scenarios the atmospheres of these massive planets are characterised by mostly solar abundances (see Fig. \ref{fig:seedImp}), with only Jupiter-like target showing limited super-solar enrichments. 

These results highlight how, in the HDM scenarios, the atmospheric enrichment of possible inner giant planets can be so marked that the resulting collisional contamination can significantly alter or overwrite the original compositional signatures of their formation process. In the LDM scenarios, on the other hand, the compositional signatures left by the planet formation process on giant planets appear to be preserved. 

\subsection{Collisional enrichment and contamination of the known giant planets}
\label{sec:impacts}
\begin{table}
	\centering
	\caption{Heavy metal collisional enrichment of the bulk envelope and atmosphere for the planets in the outer disk. Note that the outermost planet, e, underwent no collisions in the time frames considered here (last 1\,Myr and 1.16\,Myr).}
	\label{tab:enrichment}
	\resizebox{\columnwidth}{!}{\begin{tabular}{c|ccc|ccc} 
		\hline
		 \textbf{Disk-Planet}  & \multicolumn{3}{c|}{\textbf{Bulk Envelope Enrichment}} & \multicolumn{3}{c}{\textbf{Atmospheric Enrichment} } \\
        \textbf{scenario}   & Planet b & Planet c & Planet d & Planet b & Planet c & Planet d \\
       
       \hline
        HDM-HPM & 108\% & 79\% & 35\% & 31\% & 22\% &  2\% \\
        HDM-LPM &  90\% & 94\% & 41\% & 37\% & 51\& & 12\% \\
        LDM-HPM &  19\% & 20\% & 13\% & 10\% &  4\% & <1\% \\
        LDM-LPM &  17\% & 12\% & 10\% &  7\% & 11\% &  3\% \\
        \hline
        
	\end{tabular}}
\end{table}
In parallel to the enrichment and contamination of possible undiscovered inner planets, we investigated the effects the excited planetesimals have on the very giant planets embedded in the gaps of HD\,163296's disk that are at the origin of their dynamical excitation. Across our simulations, the first three planets (b, c \& d) are hit by a flux of impactors ranging from a few hundred to a couple of thousand test particles. The impacting particles originated at distances spanning the locations of the H$_2$S, CH$_4$, CO and N$_2$ snowlines, located at semimajor distances of about 19, 47, 61, and 71\,au for the reconstructed temperature profile of HD\,163296's disk midplane (see Fig. \ref{fig:chem}), and enrich the forming giant planets with heavy elements. 

The resulting compositional enrichment is significant in the HDM scenarios, where the total enrichment can reach values as high as 5 times the solar metallicity, with as much as 11\,M$_{\oplus}$ of heavy elements being added to a single planet. In the LDM scenarios, on the other hand, none of the planets accretes more that the equivalent of an Earth mass of planetesimals, meaning that the enrichment effect is negligible when assuming a solar metallicity for HD\,163296's host star and its disk. Planet e suffers no impact in all scenarios, as its faster formation timescale and larger mass make it more efficient in scattering the planetesimals that undergo close encounters with it than in accreting them.

Since the planetesimal impact rate is not constant during the growth process of the giant planets but peaks at the beginning of the runaway gas accretion process \citep{podolak2020,Turrini2021}, the previous bulk enrichments may not be reflected in the atmospheric composition of the planets unless their envelopes are well mixed. We therefore took advantage of the information on the times of the impacts recorded by our simulations to quantify the atmospheric contamination due to late accretion, to explore the scenario of inhomogeneous planetary envelopes. We focused on the contamination affecting the outermost 20\% of the planetary mass, assumed to represent the molecular gas shell \citep{Turrini2015}, hence on the impacts occurring after 0.16\,Myr since the onset of the runaway gas accretion (or, equivalently, 1.16\,Myr since the beginning of the growth process in our simulations). 

We once again find that in the HDM scenarios the atmosphere of the giant planets can be enriched by as much as 2-3 M$_\oplus$ of heavy elements, resulting in their super-stellar metallicities. In the LDM scenarios the atmospheric contamination by planetesimal impacts always results in the delivery of less than 1\,M$_\oplus$ of heavy elements and therefore produces negligible effects. In Table \ref{tab:enrichment} we summarise the above results for both the bulk envelope and atmospheric enrichment cases.

The above results, together with those discussed in Sect. \ref{sec:enrichment}, highlight how the interplay between the formation process of multi-planet systems and the protoplanetary disks in which they are embedded can play a marked role in altering the primordial compositional signatures of the native environments in the planetary atmospheres, even in absence of orbital migration, and how collisional contamination is an important process in shaping multi-planet systems born in massive disks.

\subsection{Ejection as interstellar objects}\label{sec:ejection}

As discussed in Section \ref{methods}, we considered as ejected from HD\,163296's system all planetesimals whose orbits become unbound to the host star, i.e. their eccentricity is equal to one or greater. We also estimated the fraction of planetesimals that are scattered on high eccentricity orbits with semimajor axis beyond 260\,au, i.e. beyond the orbit of the outermost planet. As shown in Table \ref{tab:ejected}, we find that significant fractions of the planetesimals originally present within HD\,163296's protoplanetary disk are ejected into interstellar space, confirming the findings of \citet{Turrini2019}. About half as many planetesimals as those ejected into interstellar space are scattered into an extended disk beyond the orbit of the fourth giant planets (see Table \ref{tab:ejected}), similar to the so-called scattered disk in the trans-neptunian region of the Solar System \citep[e.g.][]{Morbidelli2008}.

\begin{table}
	\centering
	\caption{Fraction of planetesimals within HD\,163296's circumstellar disk that are ejected as interstellar objects or are scattered in the outer disk (260-10$^4$\,au) in the different scenarios.}
	\label{tab:ejected}
	\begin{tabular}{c|cc|cc}
		\hline
		       & \multicolumn{2}{c|}{High Disk Mass} & \multicolumn{2}{c}{Low Disk Mass} \\
        
              & HPM & LPM & HPM & LPM \\
        \hline
        Disk Mass & \multicolumn{2}{c|}{0.215$\,$M$_\oplus$} & \multicolumn{2}{c}{0.05$\,$M$_\oplus$} \\
        Scattered Disk & 15\% & 6\% & 11\% & 4\% \\
        \hline
         Ejected from Disk & 25\% & 13\% & 24\% & 11\% \\
        \hline
	\end{tabular}
\end{table}

The key factor in controlling the ejection efficiency is the mass of the planets (see Table \ref{tab:ejected}): in the HPM scenarios about one fourth of the planetesimals initially present in the disk are ejected, while in the LPM scenarios this value drops to about one tenth. The disk mass has a smaller impact on the ejection efficiency driven by the exciting effect of the disk gravity, which proves dominant with respect to the damping effect of gas drag. Specifically, the constructive interplay between the planetary perturbations and the disk gravity leads to slightly higher ejection efficiencies in the high disk mass scenarios (see Table \ref{tab:ejected}). The impact of the disk gravity is more pronounced in the LPM scenarios, where the fractional increase in ejection efficiency when moving from the low disk mass to the high disk mass is four times larger than in the case of the HPM scenarios (18\% instead of 4\%). 

\begin{figure*}
	\includegraphics[width=\textwidth]{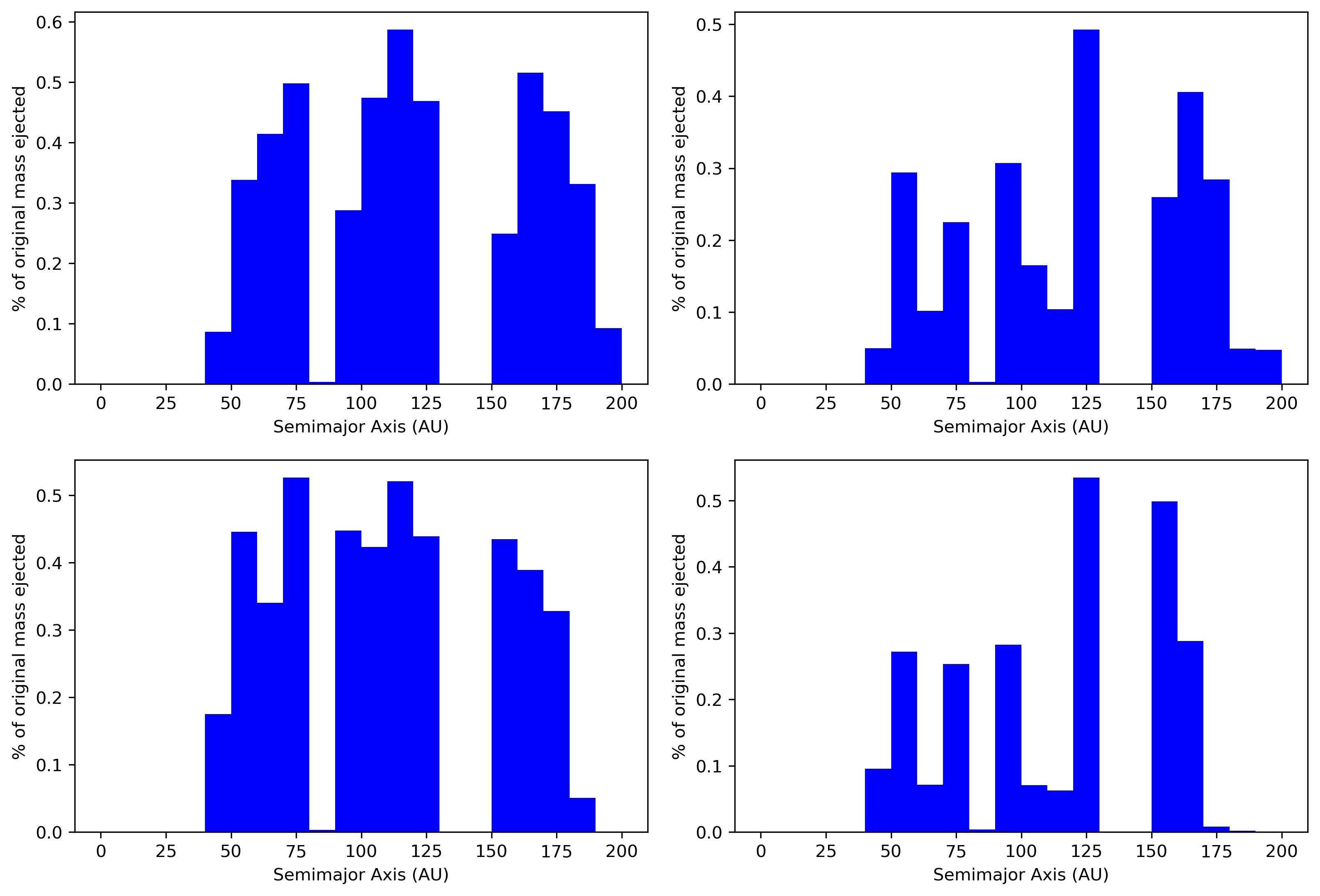}
    \caption{Planetesimal ejection efficiency across the outer disk in the four disk-planets scenarios. \textbf{Left column:} HPM scenarios. \textbf{Right column:} LPM scenarios. \textbf{Top row:} HDM scenarios. \textbf{Bottom row:} LDM scenarios. The ejection efficiency is expressed as the fraction of planetesimals originally present in each 10\,au-wide ring that end up being ejected from HD\,163296's disk.}\label{fig:ejfig}
\end{figure*}

In Figure \ref{fig:ejfig} we show how the ejection efficiency and the source regions of the ejected planetesimals vary as a function of the disk and planetary masses. The ejection efficiency is expressed as the fraction of the planetesimals originally contained into each 10\,au-wide ring that is ejected into interstellar space. The regions experiencing the highest ejection rates for all  combinations of disk and planetary masses are those between the second and third and between the third and fourth giant planets. The ring of planetesimals between the first and the second planets is significantly less depleted in the HPM scenarios than in the LPM ones (see Figure \ref{fig:ejfig}).

Under the assumptions adopted for the computation of the mass of the planetesimal disks discussed in Section \ref{methods}, we find that the LDM scenarios can eject into the interstellar space between 17 and 37\,M$_{\oplus}$. In the HDM, on the other hand, we see a factor of 5 increase in the ejected mass, that ranges between 88 and 168\,M$_{\oplus}$. Note, however, that these values provide only general indications of the efficiency of the process in producing interstellar objects, as we do not know the original mass and distribution of solids nor the planetesimal formation efficiency across the disk. The comparison between Figs. \ref{fig:chem} and \ref{fig:ejfig} points to the majority of ejected planetesimals being N-rich with an almost stellar composition with the remaining ones being rich in C and O, making them as rich in volatiles and organics or even richer than the most pristine comets observed in the Solar System \citep[see][and references therein]{Mumma2011,Altwegg2019}.

\section{Discussion}\label{sec:discussion}

In this section we first discuss the caveats on our results due to the model assumptions, followed by the physical processes and effects that, while not modelled directly, expand the impact of the dynamical and collisional processes we simulated for the compositional setup of HD\,163296's protoplanetary disk and its planets.

\subsection{Model assumptions and caveats}

In this work we assumed a planetesimal disk whose extension matches that of the observed dust disk and characterised by uniform planetesimal formation efficiency. We further assumed that the planetesimal disk was already present at the beginning of the giant planet formation process. Finally, we did not account for the enrichment of the disk gas in heavy elements due to the sublimation of ices from the inward drifting pebbles possibly supporting the core growth process.

While the assumption of matching extensions of the planetesimal and dust disks is compatible with constraints from the Solar System \citep{Kretke2012}, the radial dependence of the formation efficiency of planetesimals across the disk is currently largely unknown. Markedly lower planetesimal formation efficiencies in the outer disk regions with respect to the inner disk regions would result in limited or negligible impact of the inward transport and enrichment processes. It would also result in a sparser scattered disk and a smaller production of interstellar objects. However, the large dust abundances observed inward of the inner giant planet \citep{Isella2016,guidi2022} and their proposed origin as second-generation dust produced by high-velocity planetesimal collisions argue against widely different planetesimal formation efficiencies between inner and outer disk regions.

The presence of significant amounts of planetesimals in the disk at the onset of giant planet formation is consistent with recent results on the early conversion of dust into planetesimals during the Class 0/I stages \citep{cridland2022}. As we discuss in Sect. \ref{sec:pebbles}, our results remain qualitatively valid also in case of a prolonged planetesimal formation process throughout the giant planet formation process. The main impact on our results in such a scenario is limited to the temporal evolution of the transport and enrichment processes.

The effects of the pebble flux on the disk gas metallicity and the planetary compositions would mainly influence our results on the enrichment of the giant planets present in the disk. Specifically, the enhanced gas metallicity produced by the sublimation of the ices from the drifting pebbles would translate in an higher metallicity of the envelopes of the giant planets (which we assumed to be characterised by solar metallicity) and lower enrichments by planetesimal accretion. However, for typical disk parameters \citep{rosotti2023} the enrichment of the disk gas mainly occurs in a limited radial region inward of the snowlines \citep{BoothR2019}. This means that this process would be relevant only for the innermost giant planet at 50\,au and possible undiscovered giant planets close to the snowlines in the inner disk (see Fig. \ref{fig:chem}). In the case of solid planets, the only impact on our results would be the reduction of the enrichment by a factor equal to the ratio between the mass in pebbles and the total mass of solids (pebbles and planetesimals) in the disk.

\subsection{Interplay between dynamical transport and geophysical evolution of the planetesimals}\label{sec:differentiation}

In modelling the composition of the planetesimals and discussing the implications of their dynamical transport for the protoplanetary disk, we implicitly assumed that all planetesimals preserved their original composition, set by their formation regions in the disk midplane, and their initial budget of volatiles and organics. The meteoritic data, however, reveal that the Solar System experienced the formation of multiple generations of planetesimals and that the earliest generations experienced extensive melting and volatile loss during their differentiation and geophysical evolution \citep[e.g.][and references therein]{scott2007,lichtenberg2023}. Furthermore, meteoritic data also seem to indicate that the earliest generations of planetesimals formed in the inner Solar System, with planetesimal formation in the outer Solar System having being delayed by about 1\,Myr \citep{lichtenberg2023}.

In parallel, observational data on the evolution of the dust abundances in disks across the Class 0/I and Class II phases \citep{tychoniec2020,mulders2021,Bernabo2022} and planetesimal formation studies \citep{cridland2022} are pushing the beginning of the planet formation process closer to the beginning of the star formation process. The combination of these indications suggests that the planetesimals present in HD\,163296's inner disk may have experienced various degrees of volatile and organics loss with respect to their counterparts in the outer disk. In such a scenario, the impact of the dynamical injection and implantation of outer disk planetesimals for enriching HD\,163296's inner disk in N, C and O would be significantly enhanced with respect to what is discussed in Sect. \ref{sec:results}.

\subsection{Impact of dynamical transport on HD\,163296's habitable zone}
\label{sec:habitable}

Carbon, oxygen and nitrogen are three of the most important heavy elements representing the building blocks of life \citep[e.g.][]{herbort2024}. Because of their different volatility these elements will be present with different abundances in the planetesimals populating HD\,163296's inner disk. Specifically, carbon will be more depleted than oxygen and nitrogen will be more depleted than carbon, with nitrogen being significantly depleted in solids up to the N$_2$ snowline \citep{oberg2019,Oberg2021,Turrini2023}. As we showed in Sects. \ref{sec:implantation} and \ref{sec:injection}, the inward transport of outer disk planetesimals enriches the inner disk in all three elements. 

This enrichment in astrobiologically-important elements  can play a significant role in shaping HD\,163296's habitable zone, where the probability that planetesimals experienced major volatile loss during their geophysical evolution is high (see Sect. \ref{sec:differentiation} and \citealt{lichtenberg2023} and references therein). A simple calculation of HD\,163296's habitable zone \citep{Kasting1993, Kopparapu2013} using the current luminosity of the star (16\,L$_\odot$; \citealt{Millan-Gabet2016}) gives a ring spanning from approximately 3.8\,au to 5.5\,au. Considering the physical properties of a Main Sequence star with the same mass and metallicity (solar) as HD\,163296, the habitability zone can expand to around 9\,au (see Figure 6 from \cite{Danchi2013}). 

In two of our simulated scenarios (HDM--HPM and LDM--HPM) we find that as much as 1.3\% of the mass contained in the innermost ring (10-0.1\,au) is made up of planetesimals rich in N, C and O that formed at distances greater than 40\,au. Even if we assume the total loss of volatiles by the planetesimals originally formed in the habitable zone, the enrichment in volatile elements in this region proves greater than that estimated for the Earth and the asteroid belt in terms of water \citep[a few 10$^{-4}$, see][and references therein]{Morbidelli2000,Turrini2014,turrini2018}. Based on the study of the Earth, the enrichment in N is particularly important in light of its low solubility into magma oceans \citep{Kurokawa2022-N}. Specifically, \citet{Kurokawa2022-N} argue that, if the Earth's current N budget is primordial and inherited from its formation process, Earth's initial abundance of N should have been an order of magnitude higher. The delivery of N-rich planetesimals to the habitable zone could therefore support the formation of N-dominated atmospheres like that of the Earth.

Finally, as shown in Fig. \ref{fig:chem} the inward transport of planetesimals can also affect planetesimals enriched in sulphur by the condensation of H$_2$S, but data from meteorites \citep{lodders2010,palme2014} and protoplanetary disks \citep{Kama2019,Riviere-Marichalar2020,Riviere-Marichalar2021,LeGal2021} suggest that most S is incorporated into refractory materials close to the host star. As a result, the enrichment in S due to the dynamical transport of planetesimals is plausibly limited.

\subsection{Extended impact flux and dynamical transport in the case of a pebble-rich disk}\label{sec:pebbles}

While the focus of this study is on the planetesimals, the process of dynamical transport we discuss applies also to the case of dust-dominated or pebble-dominated scenarios for HD\,163296's disk.  Specifically, when the growing planets reach the so-called pebble isolation mass, dense rings of pebbles accumulate at the pressure bumps outside each planet orbit \citep{Morbidelli2012, Lambrechts2014}. These pebble-dense rings provide favourable sites for forming planetesimals by streaming instability \citep{johansen2017} and can continue to grow in mass until an outer planet reaches its pebble isolation mass, blocking the incoming flux of pebbles, or the gas begins to dissipate typically within 10 to 20\,Myr \citep{Hernandez2007, Fedele2010, Ribas2014}.

Once planetesimals form in these rings, they get dynamically excited and undergo the same orbital evolution of the dynamical tracers in our simulations \citep{Eriksson2021}. This means that, if HD\,163296's disk has been characterised by a continuous process of planetesimal formation in its outer regions (as proposed also for the outer Solar System, \citealt{lichtenberg2023,Neumann2024}), the dynamical injection of the planetesimals and the collisional contamination of the planets embedded in the disk can evenly spread across the whole life of the system instead of reaching their peak intensity when the giant planets undergo their runaway gas accretion. This scenario likely applies to the two outermost rings, as the dust production by planetesimal impacts in these orbital regions does not appear to be  efficient enough to explain the local dust solely as second-generation dust \citep{Turrini2019}.

Furthermore, if pebbles are still present in the system when the gaseous disk dissipates an additional phase of dynamical evolution occurs. Once the gas has dissipated, the pebbles do not feel the pressure bump and the damping effects of the gas any more. As a consequence, the pebble rings are quickly excited by the giant planets and the pebbles start to behave dynamically like planetesimals. An additional flux of ice-rich pebbles is expected, which would reach the inner regions of the planetary system like discussed in Sect. \ref{sec:results}. This third wave of contaminating bodies is expected to extend in time the impacting flux on existing planets, increasing their budget of carbon, oxygen and nitrogen. 

Differently from planetesimals, however, due to the small sizes of the pebbles the ices and organics they contain will undergo prompt sublimation due to the higher temperatures they will experience during their inward transport. The capability of this pebble flux to transport volatiles and organics to the inner disk will therefore depend on the ratio between the time they need to reach thermal and chemical equilibrium with the surrounding environment and their flight time before being accreted. This late flux of pebbles will likely be more effective in contaminating the observed giant planets, which will be closer to their source regions, than any possible inner planet.

\subsection{Interplay between dynamical transport and pebble accretion in the inner disk}

As discussed by \citet{Turrini2019}, the dynamical excitation and transport of the planetesimals will result in their super-sonic motion with respect to the gas, which in turn will result in their heating and thermal ablation \citep{Tanaka2013}. In their study focused on the Solar System and HL Tau's disk, \citet{Eriksson2021} argue that the ablation of scattered planetesimals may replenish the pebble population in the inner disk regions, supporting the process of planetary growth by pebble accretion by local planetary bodies \citep{johansen2017}. 

In parallel, \citet{Turrini2019} showed how the collisional cascade triggered by excited planetesimals can produce tens of Earth masses of dust and explain the abundant dust population observed inward of HD\,163296's first giant planet \citep{Isella2016}. The study of the ejecta size distribution after Hayabusa 2's impact experiment on asteroid Ryugu point to the ejected dust particles ranging in size from one mm to several decimetres, with a characteristic size of about one cm \citep{Wada2021}. 

Collisional debris and second-generation dust sharing a similar size distribution will behave dynamically like pebbles and also support the planetary growth process by pebble accretion \citep{Turrini2023} in the inner disk of HD\,163296, overcoming the obstacle posed to the inward drift of pebbles from beyond 40\,au by the barrier effect of the four giant planets. As a result, the dynamical excitation and inward transport of the outer disk planetesimals by the giant planets can support the formation of inner undiscovered planets, like those discussed in Sect. \ref{sec:enrichment}, and more generally extend the duration of the pebble accretion and streaming instability processes in the inner disk.

\section{Conclusions}\label{sec:conclusions}

Over the past decade a growing body of studies has been arguing how the appearance of giant planets in protoplanetary disks naturally triggers an intense phase of dynamical excitation of the surrounding planetesimal disk, resulting in the dynamical and collisional transport of volatile elements across different orbital regions and planetary bodies \citep{Turrini2011,Turrini2014,Turrini2015,turrini2018,Turrini2018b,Turrini2019,Raymond2017,ronnet2018,pirani2019,Bernabo2022,Seligman2022,shibata2022,SainsburyMartinez2024}. In parallel, observational and laboratory studies have been providing more and more evidence supporting the importance of these processes in shaping the early history of the Solar System \citep{Sarafian2013,Sarafian2014,Sarafian2017,DeSanctis2015,bischoff2018,Trigo2019SS,Nittler2019,Kebukawa2020,Kurokawa2022-N,Taylor2024}. 

In this work we used as our case study the HD\,163296 system, one of the best characterised protoplanetary disks \citep{Isella2016,isella2018,BoothA2019,Kama2020,Wichittanakom2020} that has been suggested to be host to four or more giant planets \citep{Isella2016,Liu2018,Pinte2018,guidi2018,zhang2018,Teague2018,Teague2019,mesa2019,izquierdo2022,alarcon2022,garrido-deutelmoser2023} and a dynamically and collisionally excited planetesimal disk \citep{Turrini2019}. Our goal is to investigate the impact of the dynamical and collisional transport processes in infant planetary systems with giant planets on wide orbits like those revealed by ALMA surveys of protoplanetary disks \citep[e.g.][]{ALMA2015,Isella2016,fedele2017,Long2018,Andrews2018}.

We explored four end-member scenarios that span the possible combinations of the estimated disk and planetary masses and allow us to assess the impact of these physical parameters on the dynamical evolution of the system as a whole. We find that the appearance of the giant planets in the host disk always results in large-scale dynamical excitation of the planetesimals. The intensity of the dynamical excitation process proves to be controlled by the masses of the giant planets yet we find that the disk mass has smaller but non-negligible effects, with the excitation effect of the disk gas proving more important than the damping one of gas drag in the outer disk regions. The dynamical excitation of the planetesimal disk results in the creation of a massive population of exo-comets that transports volatile elements to the inner disk regions, enriching existing planets and their atmospheres in these astrobiologically important elements, and produces a large population of interstellar planetary objects.

Specifically, we find that:
\begin{enumerate}
    \item The giant planets in the outer disk deliver planetesimals to the inner disk both through dynamical injection and implantation. Injected planetesimals periodically cross the inner disk on eccentric orbits, while implanted planetesimals become decoupled from the outer disk to permanently reside in the inner disk. HD\,163296's giant planets between 50 and 260\,au are capable of delivering large amounts of planetesimals inward of 40\,au.
    \item Dynamical implantation and injection influence different but overlapping regions of the inner disk. Implantation is more efficient between 10 and 30\,au, while injection is more efficient between 20 and 40\,au. Implantation can increase the local mass of the affected ring up to $\approx$10\%, while injection can temporarily increase it up to $\approx$50\%. These values are comparable or higher than those characterising the same process in the Solar System \citep{Turrini2014,Raymond2017,ronnet2018} but the affected area is an order of magnitude larger.
    \item Higher planetary masses result in excited planetesimals penetrating deeper into the inner disk and allow for their implantation in the innermost 10\,au. Higher disk masses enhance the excitation in the outer disk but the stronger gas drag they exert circularises the orbits farther away from the star, favouring the region between 20 and 30\,au in HD\,163296's case.
    \item The dynamical transport process results in the collisional enrichment of possible planets populating the inner disk in the volatile elements C, O and N (and, possibly, S). For higher disk masses, solid planets spanning from lunar to super-Earth masses and residing beyond 10\,au are enriched to higher or comparable levels than the Earth, while for lower disk masses the enrichment is on average an order of magnitude smaller. Giant planets ranging from sub-Neptunian to Jovian masses and residing in the same orbital region can have their atmospheres enriched from two to nine times the stellar metallicity proportionally to their mass. 
    \item The dynamical transport process can affect also the very giant planets that trigger it. For the higher disk masses, planetesimal accretion significantly enhances the bulk metallicity of planets b and c and moderately that of planet d. The atmospheric metallicity of these planets is increased moderately to limitedly, respectively. For the lower disk masses, the impact of planetesimal accretion on the planetary metallicity is generally negligible. Due to our assumptions on its faster formation, planet e is never affected by planetesimal accretion.
    \item The dynamical excitation of the planetesimals scatters a significant fraction of them (up to 40\%) beyond the orbit of the outermost planet. In all simulated cases, about one third of the scattered planetesimals remain bound to HD\,163296 forming an extended and excited planetesimal belt as in the case of the Solar System.  The remaining two thirds are instead ejected into interstellar space, making systems like HD\,163296 efficient factories of interstellar objects. 
\end{enumerate}    
Building on the above findings, we can draw the following global implications for the evolution of protoplanetary disks hosting young massive planets:
\begin{enumerate}
    \item Depending on the specific architecture of the system, the dynamical and collisional transport process can influence the composition of already formed planets, including those in the habitable zone under the right conditions. In the case of solid planets, this can have a major impact on their budget of volatiles and astrobiological material. In the case of giant planets, the same processes can alter or overwrite their original atmospheric compositional signatures.
    \item The interplay between dynamical transport, thermal ablation and collisions can revert significant fraction of the injected and implanted planetesimals back into dust and pebbles. This can prolong the temporal window available to form planets by pebble accretion, bypassing the barrier effect of existing massive planets to the inward drift of the pebbles from the outer disk.
    \item The dependence of the collisional enrichment on the disk mass can, in principle, put further constraints on the disk mass. Specifically, signatures of planetesimal impacts, like the presence of dust or refractory elements in the atmospheres of giant planets, would point to high disk masses while their absence would favour lower disk masses. 
\end{enumerate}
Overall the picture depicted by our results indicates that studies aimed to constrain the formation history of planets through their atmospheric characterisation need to account for the global architecture of the host planetary system, as the presence of massive planets can imply that the present day atmospheric signatures are not a reflection of the primordial ones.

\begin{acknowledgements}

We thank the reviewer for their comments that helped give more clarity to the manuscript. 
This work is supported by the Fondazione ICSC, Spoke 3 “Astrophysics and Cosmos Observations'', National Recovery and Resilience Plan (Piano Nazionale di Ripresa e Resilienza, PNRR) Project ID CN\_00000013 “Italian Research Center on High-Performance Computing, Big Data and Quantum Computing'' funded by MUR Missione 4 Componente 2 Investimento 1.4: Potenziamento strutture di ricerca e creazione di “campioni nazionali di R\&S (M4C2-19)'' - Next Generation EU (NGEU). D.P. and P.S. acknowledge the support from the Istituto Nazionale di Oceanografia e Geofisica Sperimentale (OGS) and CINECA through the program “HPC-TRES (High Performance Computing Training and Research for Earth Sciences)” award numbers 2022-05 and 2022-02. The authors acknowledge the support from the ASI-INAF grant no. 2021-5-HH.0 plus addenda no. 2021-5-HH.1-2022 and 2021-5-HH.2-2024 and grant no. 2016-23-H.0 plus addendum no. 2016-23-H.2-2021 and from the PRIN INAF 2019 PLATEA and HOT-ATMOS and the INAF Main Stream project “Ariel and the astrochemical link between circumstellar discs and planets” (CUP: C54I19000700005). This work was partly supported by the Italian Ministero dell’Istruzione, Università e Ricerca through the grant Progetti Premiali 2012-iALMA (CUP C52I13000140001). This project has received funding from the European Union’s Horizon 2020 research and innovation program under the Marie Sklodowska--Curie grant agreement No. 823823 (DUSTBUSTERS) and from the European Research Council (ERC) via the ERC Synergy Grant ECOGAL (grant 855130). JMT-R acknowledges financial support from the project PID2021-128062NB-I00 funded by MCIN/AEI/10.13039/501100011033. This work has made use of data from the European Space Agency (ESA) mission {\it Gaia} (\url{https://www.cosmos.esa.int/gaia}), processed by the {\it Gaia} Data Processing and Analysis Consortium (DPAC, \url{https://www.cosmos.esa.int/web/gaia/dpac/consortium}). Funding for the DPAC has been provided by national institutions, in particular the institutions participating in the {\it Gaia} Multilateral Agreement. This research has made extensive use of NASA’s Astrophysics Data System, funded by NASA under Cooperative Agreement 80NSSC21M00561. The authors would like to acknowledge the computational support from John Scige Liu and the Genesis cluster at INAF-IAPS.
\end{acknowledgements}

\bibliographystyle{aa}
\bibliography{Bibliography}

\end{document}